\documentclass[journal]{IEEEtran}
\usepackage{cite}
\usepackage{amsmath,amssymb,amsfonts}
\usepackage{algorithmic}
\usepackage{graphicx}
\usepackage{textcomp}
\usepackage{xcolor}
\usepackage{booktabs}
\usepackage{array}
\usepackage{hyperref}
\usepackage{comment}
\hypersetup{
    colorlinks=true,
    linkcolor=black,
    citecolor=black,
    urlcolor=blue
}

\begin{document}



\title{Pre- and Post-Exercise Monitoring of Physiological Skin States in Sportsmen Using 77 GHz FMCW Radar and Wavelet Scattering Transform \vspace{-5pt}}


\author{Vijith Varma Kotte,~\IEEEmembership{Member,~IEEE,}
Muhammad Mahboob Ur Rahman,~\IEEEmembership{Senior Member,~IEEE,}
Sajid Ahmed,~\IEEEmembership{Senior Member,~IEEE,}
and~Tareq Y. Al-Naffouri,~\IEEEmembership{Fellow,~IEEE}%
\thanks{The authors are with the Department of Electrical and Computer Engineering, CEMSE Division, King Abdullah University of Science and Technology (KAUST), Thuwal 23955, Saudi Arabia. Corresponding author: Muhammad Mahboob Ur Rahman (email: muhammad.rahman@kaust.edu.sa). This research was supported by funding from KAUST Center of Excellence for Smart Health (KCSH), under award number 5932.}
\vspace{-25pt}}
\maketitle

\begin{abstract}


This paper presents a contactless radar sensing framework that classifies physiologically-induced changes in the superficial skin layer following intense fixed-duration physical activity along with water intake restriction. A 77 GHz frequency-modulated continuous wave (FMCW) radar is used to capture high-resolution range profiles from the subject's chest, seated in front of the radar, before and after sports activity. Since the electromagnetic penetration into biological tissue is inherently shallow at millimeter-wave frequencies, the sensing is confined primarily to the stratum corneum and upper epidermis. We demonstrate that exercise-induced thermoregulation, perspiration residue deposition, and transient shifts in superficial tissue hydration produce measurable alterations in quasi-static radar reflectivity. We utilize single-chirp snapshots of duration 102.4 $\mu$s in order to suppress confounders such as slow-varying cardiopulmonary motion, followed by lightweight feature extraction via the Wavelet Scattering Transform (WST) and classification using light-weight machine learning models suitable for edge nodes. Leave-one-subject-out cross-validation on a dataset of 15 sportsmen measured pre- and post-exercise yields a mean classification accuracy of 91.55\% with a 95\% confidence interval of [84.27, 98.83]\%. Interpretability analysis reveals that low-order scattering coefficients (capturing overall signal energy and amplitude stability) dominate discriminative power, consistent with quasi-static changes in skin surface permittivity. The results establish a proof-of-concept for radar-based, edge deployable classification of post-exercise skin states, opening new avenues for non-contact physiological monitoring in sports science.

\end{abstract}
%
\vspace{-8pt}

\section{Introduction}


Intense physical exercise combined with water intake restriction is a common experimental paradigm used to assess the physiological well-being, endurance, and recovery capacity of athletes and sportsmen under thermal and hydration stress \cite{periard2021exercise}. Such conditions induce significant systemic stress, which, among other changes, triggers measurable changes in the physiological state of the superficial skin. These changes include vasodilation, hyperthermia, eccrine sweat secretion, perspiration residue deposition, and dynamic shifts in stratum corneum water content \cite{periard2021exercise}. Consequently, the skin—particularly the skin-air interface and upper epidermal layers—undergoes alterations in its dielectric and morphological properties.

Measuring the exercise-induced skin state changes via internet of things (IoT) devices is important due to a number of reasons. It enables non-invasive hydration monitoring, early detection of heat stress or exhaustion, personalized recovery guidance, objective workout intensity estimation, longitudinal skin health tracking, and seamless integration with smart IoT environments for automated interventions (e.g., cooling, hydration alerts) \cite{periard2021exercise,liaqat2022personalized}. However, traditional assessment methods rely on invasive biomarkers or wearable sensors that require skin contact, frequent calibration, and active user compliance, limiting scalability in smart healthcare and athletic training settings \cite{siyoucef2025internet, posada2019mild}.

Millimeter-wave radar, particularly at 77 GHz, offers a natural IoT-based solution for contactless skin state measurement. At this frequency, the propagation in biological tissue is strongly attenuated due to high dielectric losses \cite{gabriel1996dielectric}. This physical property confines the radar sensing interaction primarily to the skin-air interface, stratum corneum, and upper epidermis—which are all influenced by the exercise-induced physiological changes \cite{sasaki2019optical}. Thus, 77 GHz IoT radar is inherently sensitive to variations in permittivity and reflectivity of superficial skin layers caused by sweat residue, hydration shifts, and thermoregulation. Further, when deployed at the edge, the IoT radar can perform local inference, transmit only lightweight latent data for split computing, and operate without body-worn devices, preserving user comfort and privacy.

Prior RF-IoT-based physiological monitoring solutions have mainly focused on vital-sign extraction or hydration-level classification. Continuous-wave and Doppler radars have been used to estimate respiration and heart rates using phase or micro-Doppler modulation over multiple slow-time samples \cite{Vijith13}. Other RF-IoT solutions include non-contact hydration classification using WiFi, software-defined radio, and radar measurements \cite{buttar2023noncontact,sosa2026radio}. Contact-based IoT solutions such as electrodermal and dielectric sensing have also shown that external stressors such as sweat alters physiological response of the skin (e.g., change in hydration level) \cite{liaqat2022personalized,choi2021novel,zhang2022contactless}. However, when it comes to mmWave IoT radar, claims such as dehydration monitoring require careful interpretation due to the shallow penetration depth of mmWave frequencies and cardiopulmonary motion, that can confound measurements. Thus, we are of opinion that, a lightweight, interpretable classification of skin states—rather than exact hydration estimation—is a physically and theoretically sound approach. 

Motivated by these considerations, the contributions of this work are: (i) a 77 GHz FMCW IoT radar dataset from 15 athletes under pre-activity and fluid-restricted post-activity conditions; (ii) an edge-deployable inference pipeline using single-chirp sampling (102.4~$\mu$s) to suppress cardiopulmonary signatures, combined with wavelet scattering transform features and lightweight machine learning; and (iii) a leave-one-subject-out validation achieving 91.55\% mean accuracy, demonstrating that post-exercise superficial radar-response changes are discriminative for subject-independent skin state classification.

\vspace{-3pt}

\section{IoT Radar-based Skin Sensing: Challenges \& Solutions}

\noindent\textit{Challenge 1: Ambient Skin Sensing in IoT Environments.} IoT-enabled health monitoring in smart gyms, athletic training facilities, and remote healthcare settings is traditionally done by means of wearable sensors impose adherence burdens, skin irritation, and frequent maintenance, limiting scalability. 

\textit{Solution:} Millimeter-wave IoT radar could offer viable insights into contactless skin physiological sensing, thanks to the physics guiding the radio frequency signal interaction with the human tissue. For the 77 GHz IoT radar used in this work, the wavelength is 3.9 mm, which implies that the propagation in biological tissue is strongly attenuated due to high dielectric losses. Consequently, the radar response is dominated by the skin--air interface and superficial epidermal region (stratum corneum and upper epidermis) rather than deeper tissues \cite{foster2002electromagnetic,gabriel1996dielectric}. This physical constraint guides the radar IoT-based skin state monitoring, as it isolates surface-level physiological changes without interference from deep tissue and fat layers.

\noindent\textit{Challenge 2: Capturing Exercise-Induced Changes in Superficial Skin Layers.} Intense physical activity in the heat, combined with fluid restriction triggers multiple surface-level responses—thermoregulatory vasodilation, elevated skin temperature, eccrine sweat secretion, perspiration residue deposition, and shifts in stratum corneum water content. The radar IoT sensor must be sensitive enough to detect these subtle dielectric modulations reliably. 

\textit{Solution:} The physiological stressors (i.e., heat, exercise, water intake restriction) alter the local complex permittivity $\epsilon^*$ and surface conductivity of the skin-air interface. For a simplified normal-incidence model, the reflection coefficient is $\Gamma \approx (1-\sqrt{\epsilon^*})/(1+\sqrt{\epsilon^*})$, so changes in $\epsilon^*$ directly modulate the amplitude and phase of radar returns \cite{zhang2022contactless,choi2021novel,liaqat2022personalized}. Thus, 77 GHz radar is sensitive enough to transduce exercise-induced skin state variations into measurable baseband signal perturbations, enabling reliable differentiation between pre- and post-exercise physiological skin states.

\noindent\textit{Challenge 3: Suppressing Cardiopulmonary Confounders.} A major challenge for radar-based physiological IoT sensing is distinguishing intended skin-state changes from periodic vital-sign artifacts (respiration at 0.1--0.5 Hz, cardiac activity at 0.8--2.5 Hz). Traditional approaches require slow-time sampling and complex filtering, increasing computational load on resource-constrained edge nodes. 

\textit{Solution:} Our framework samples single FMCW chirps of duration $T=102.4~\mu$s, several orders of magnitude shorter than cardiopulmonary cycles. This design explicitly suppresses explicit slow-time vital-sign signatures, emphasizing instantaneous amplitude, phase, and spectral variations of the radar response. The result is a compact, low-complexity representation suitable for lightweight machine learning on IoT edge devices, while ensuring that classification decisions are driven by genuine skin-state changes rather than artifacts due to periodic cardiopulmonary motion.

\section{Experimental Protocol and Data Acquisition}
\subsection{Configuration of IoT Radar System}
Data were acquired using an INRAS 77 GHz FMCW MIMO radarbook operating with a 2 GHz bandwidth, two transmit (Tx), and sixteen receive (Rx) antennas. The radar IoT system was configured using a single Tx--Rx channel with a standoff distance of 1 m from the subject's chest. The radar emitted linear chirps with a duration of 102.4 $\mu$s and 256 fast-time samples per chirp. This configuration provides a range resolution of approximately 7.5 cm, sufficient to separate the subject's chest response from background clutter. Fig.~1 shows the experimental setup.
\begin{figure}[!t]
    \centering
    \includegraphics[height=0.5\linewidth, width=0.8\linewidth]{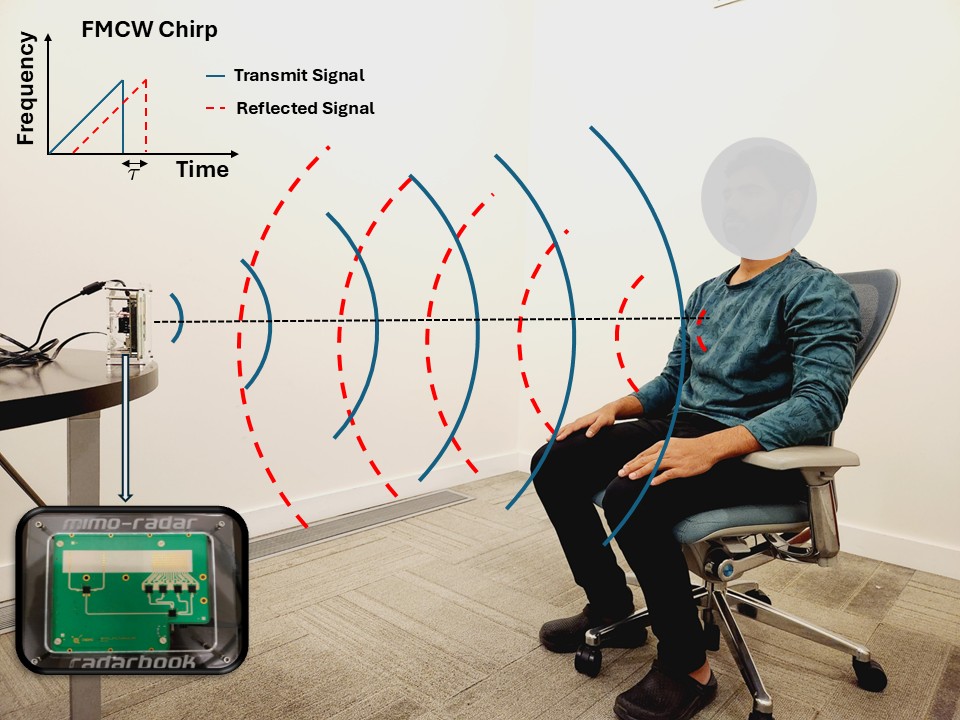}
    \caption{Experimental setup: contact-less physiological skin-state monitoring using 77 GHz FMCW radar.}
    \vspace{-10pt}
    \label{fig:DataCollection}
\end{figure}
\vspace{-10pt}
\subsection{Data Collection Protocol and Data Labeling Strategy}
Fifteen university-level ultimate frisbee athletes participated in the study under IRB approval (KAUST IBEC Protocol 24IBEC032). Written informed consent was obtained from all participants. The experimental protocol followed a controlled pre/post-exercise design:
\begin{itemize}
    \item \textit{Pre-Activity Reference State:} Radar recordings were done after subjects rested for 15 minutes in a hot outdoor environment, providing the pre-exercise reference skin state.
    \item \textit{Post-Exercise State:} Subjects engaged in 2 hours of intense sports activity without water intake. Post-exercise radar recordings were collected within 10 minutes after cessation.
\end{itemize}
Each recording session lasted 1--2 minutes. For each subject and condition, 125 single-chirp samples are selected for analysis, yielding 1875 samples per class and 3750 samples in total. Labels were assigned as pre-activity and post-exercise superficial skin states. For each selected sample, one chirp and one receive channel corresponding to the dominant chest response were used to form a $1 \times 256$ fast-time vector. The same selection procedure was applied uniformly across subjects and sessions. No multi-chirp averaging was used, preserving the single-chirp quasi-static representation adopted in this study.

\section{Methodology}
The acquired single-chirp radar samples are processed through the following pipeline: FMCW range profiling, chest-response selection, WST feature extraction, and machine-learning classification.
\vspace{-10pt}
\subsection{Signal Model}
The basic system model of the proposed system is depicted in Fig. \ref{fig:DataCollection}. The FMCW chirp signal transmitted from transmitting antenna in complex form can be expressed as: $ s(t) = e^{j2\pi (f_{c}+\frac{B t}{2T}) t } $, where $T$ is chirp duration, $f_c$ is carrier frequency, $B$ is bandwidth of the transmitted signal, $t$ is fast-time index $(0 < t < T)$. The transmitted signal gets reflected from the targets and is picked up by the receive antennas. Considering $L$ targets in the field-of-view of radar, the received signal becomes:
\begin{equation}
r(t) = \sum_{l=1}^{L} \tilde{\alpha}_l e^{j2\pi\left(f_c + \frac{B t}{2T}\right) (t - \tau_{l})} + \bar{n}(t) \label{Eq_rx}
\end{equation}
\noindent where $\tilde{\alpha}_l$ and $R_l$ are the attenuation factor and distance of $l$th target, respectively. $\tau_l = \frac{2 R_l}{c}$ is the round trip delay of the $l$th target, $c$ is the speed. The term $\bar{n}(t)\sim {\cal N}(0, \sigma^2)$ is noise.  
The received signal in \eqref{Eq_rx} is multiplied by the transmitted chirp signal and then passed it through a low-pass filter, known as demodulation. The demodulated received signal becomes
\begin{equation}
    x(t) = \sum_{l=1}^{L} \tilde{\alpha}_l e^{j2 \pi f_{c}\tau_l} e^{j2\pi\left(\frac{B\tau}{2T}\right)t} + n(t), \label{Eq_r1}
\end{equation}
where $n(t) = \bar{n}^{*}(t) e^{j2\pi (f_{c}+\frac{B t}{2T}) t}$. Substituting $\tau_l$ in \eqref{Eq_r1}, the demodulated received signal can be written as
\begin{equation}
x(t) = \sum_{l=1}^{L} \alpha_l e^{j2\pi\left(\frac{2B R_{l}}{c T}\right)t} + n(t), \quad 0 \leq t \leq T \label{Eq_r2}
\end{equation}
where $\alpha_l = \tilde{\alpha}_l e^{j\frac{4 \pi f_{c} R_{l}}{c}}$. The term $e^{j2\pi\left(\frac{2B R_{l}}{c T}\right)t}$ in \eqref{Eq_r2} is complex sinusoidal signal which contains the range information of the targets. The beat frequencies, $f_{b_l} = \frac{2B R_{l}}{c T}$, can be estimated by performing a Fast Fourier Transform (FFT) to the received signal in \eqref{Eq_r2} across the fast-time, i.e., range FFT. In this work, the range profile is used to identify the dominant chest reflection, while the corresponding single-chirp radar vector is used for WST-based feature extraction.
\vspace{-10pt}
\subsection{Pre-Processing Stage}
Each selected single-chirp vector is first windowed using a Hamming window to reduce range sidelobes:
\begin{equation}
x_w(t) = x(t) \cdot w(t)
\end{equation}
An FFT is then applied across the fast-time dimension to obtain the range profile. A second-order-section (SOS) Butterworth band-pass filter is used to retain the range interval containing the dominant chest reflection and suppress background clutter and multipath components. The resulting $1 \times 256$ radar vector is passed to the WST feature-extraction stage.

\subsection{Wavelet Scattering Transform (WST) for Feature Extraction}
The WST is a deep convolutional network with fixed, mathematically designed filters that extract low-variance, translation-invariant features from time-series data \cite{mallat2012group}. It cascades wavelet modulus operations with low-pass filtering, producing a hierarchical representation robust to small temporal shifts and amplitude scaling. We employ Gabor (analytic Morlet) wavelets with quality factor $Q$ controlling the frequency resolution.

\vspace{-10pt}
\subsection{Machine Learning Classification}
Extracted features are fed into five classifiers: Naive Bayes (NB), Support Vector Machine (SVM), K-Nearest Neighbors (KNN), Kernel-based method, and a Feedforward Neural Network (NN). Model selection prioritizes computational efficiency for potential edge deployment. Validation employs 5-fold cross-validation and Leave-One-Subject-Out cross-testing to evaluate generalization across inter-subject variability.
\vspace{-10pt}
\section{Results and Analysis}
\begin{figure}[!t]
\centering
\includegraphics[width=\columnwidth]{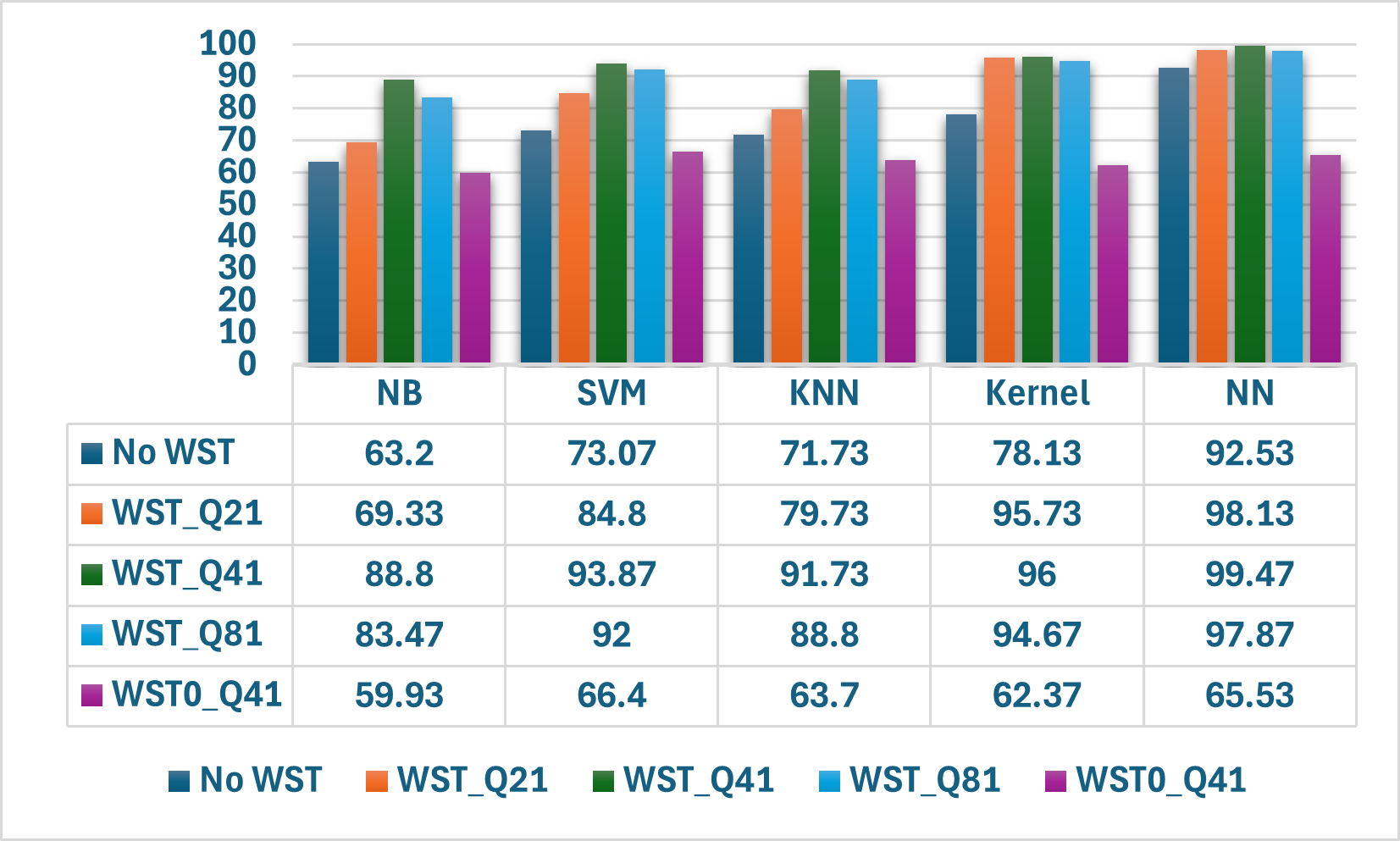}
\caption{Classification accuracy across NB, SVM, KNN, Kernel, and NN methods for different WST quality factors ($Q=[2,1], [4,1], [8,1]$). }
\vspace{-10pt}
\label{fig:accuracy}
\end{figure}
For the proposed single-chirp analysis, one chirp and one receive
channel corresponding to the dominant chest reflection are
selected, yielding a $1 \times 256$ fast-time vector per sample.
The resulting vectors are processed using a wavelet scattering
transform with analytic Morlet wavelets. The invariance
scale is set to 128 samples, corresponding to half of the signal
length. The WST output is arranged as a tensor of size
$25 \times 8 \times 3750$, where 25 denotes the number of
scattering paths, 8 denotes the number of temporal scattering
windows, and 3750 denotes the total number of radar samples.
For classifier input, this tensor is reshaped into a
$3750 \times 200$ feature matrix by concatenating the
scattering-path and temporal-window dimensions. Eventually a number of ML classifiers are evaluated using the extracted WST features.

Fig.~\ref{fig:accuracy} compares the classification
performance obtained without WST and with WST features
using different quality factors, namely $Q=[2,1]$, $Q=[4,1]$,
and $Q=[8,1]$. Across classifiers, $Q=[4,1]$ provides the best
tradeoff between feature compactness and discriminative
resolution. The WST features consistently outperform the
raw-signal baseline, indicating that the scattering representation
captures stable multiscale structure in the radar response.
The adopted reshaping strategy, denoted WST\_Q41, also
outperforms the alternative unfolding strategy WST0\_Q41,
suggesting that preserving the joint scattering-path and
temporal-window structure improves class separability.

To evaluate subject-independent generalization and avoid
subject-level data leakage, leave-one-subject-out (LOSO)
cross-validation is used as the primary validation protocol.
Random 5-fold cross-validation (CV) is repeated over 10 runs only
as a secondary reference. As summarized in Table~\ref{tab:LOSO},
the LOSO accuracies are lower than the random CV results,
as expected under a stricter subject-independent setting.
Nevertheless, the NN classifier achieves the highest mean LOSO
accuracy of $91.55\%$ with a 95\% confidence interval (CI) of
$[84.27,98.83]\%$, demonstrating that the proposed radar-WST
pipeline can discriminate pre-activity and post-exercise
superficial skin states across unseen subjects.

\begin{figure}[ht]
\centering
\includegraphics[width=\columnwidth]{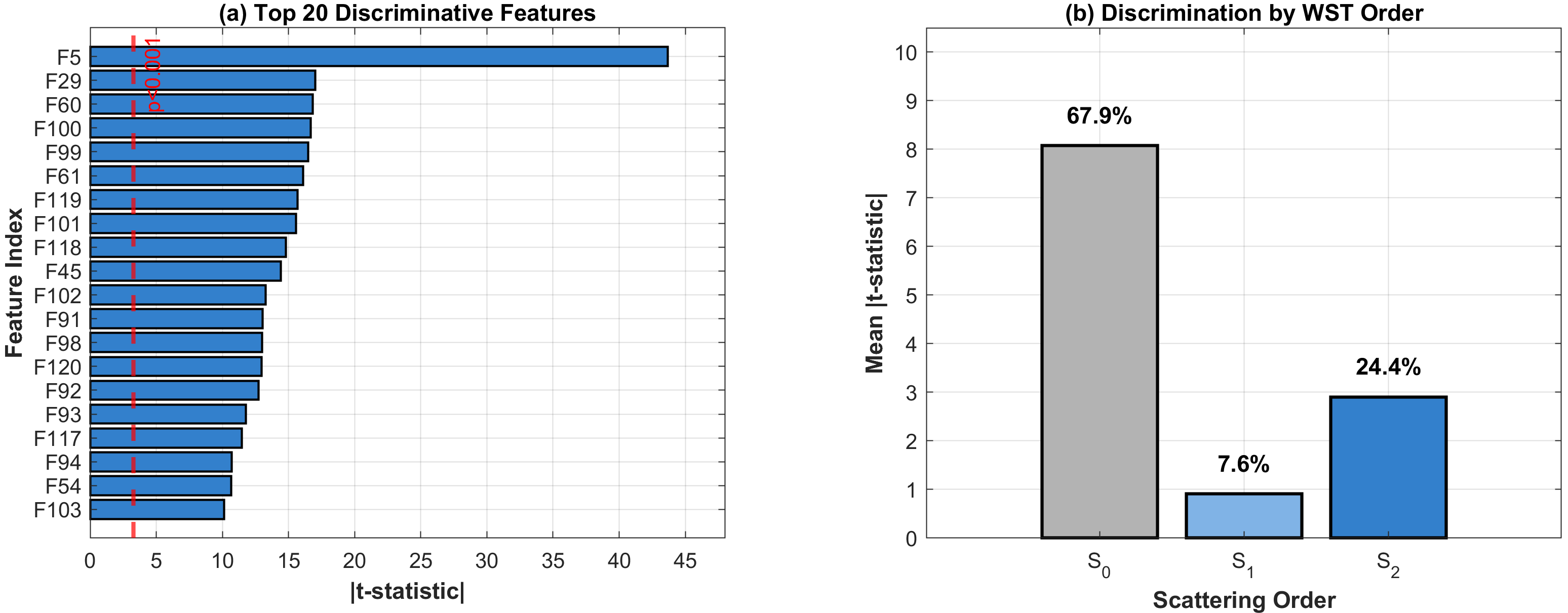}
\caption{(a) Top 20 discriminative WST coefficients ranked by $|t|$-statistic. (b) Relative contribution by scattering order: $S0$ dominates, capturing low-pass energy variations consistent with quasi-static surface reflectivity shifts.}
\vspace{-10pt}
\label{fig:feature_importance}
\end{figure}

To interpret the discriminative WST features, two-sample
$t$-statistics are computed across the 200 WST coefficients.
This analysis is used only as an exploratory feature-ranking
tool, since multiple chirps from the same subject are not fully
independent. As shown in Fig.~\ref{fig:feature_importance}, the dominant features are concentrated in low-order scattering coefficients.
Specifically, the zeroth-order coefficients contribute
approximately $67.9\%$ of the total mean absolute
$t$-statistic, followed by second-order ($24.4\%$) and
first-order ($7.6\%$) coefficients. Since zeroth-order
coefficients mainly describe low-pass signal energy and
amplitude stability, their dominance is consistent with
quasi-static changes in the radar reflection from the superficial
skin interface after exercise.
\begin{table}[h] 
\centering 
\caption{{Mean classification accuracies along with standard deviation (SD) of various classifiers using 10-run CV and LOSO testing.}}
\label{tab:LOSO}
\begin{tabular}{|p{0.11\linewidth} | p{0.24\linewidth}| p{0.24\linewidth} | p{0.2\linewidth} |} 
\hline 
\textbf{Classifier} & \textbf{CV Test Accu. $\pm$ SD (10 runs)} & \textbf{LOSO Test Accu. $\pm$ SD (\%)} & \textbf{95\% CI (LOSO)} \\ 
\hline 
NB & 89.84 $\pm$ 1.52 & 76.35 $\pm$ 15.98 & [68.26, 84.43]\\ 
SVM  & 92.16 $\pm$ 1.12 & 80.32 $\pm$ 15.85 & [72.30, 88.34] \\ 
KNN  & 90.29 $\pm$ 0.87 & 77.87 $\pm$ 14.55 & [70.50, 85.23] \\ 
Kernel & 96.51 $\pm$ 1.11 & 85.07 $\pm$ 13.90 & [78.03, 92.10] \\ 
NN & \textbf{99.28 $\pm$ 0.56} & \textbf{91.55} $\pm$ 14.39 & \textbf{[84.27, 98.83]} \\ 
\hline 
\end{tabular} 
\vspace{-10pt}
\end{table}

\vspace{-10pt}

\section{Conclusion}

This study demonstrated an IoT-based contactless 77 GHz FMCW radar framework for classifying pre- and post-exercise superficial skin states across 15 athletes. The integration of single-chirp radar sensing, WST-based feature extraction, and lightweight ML classifiers achieves 91.55\% LOSO accuracy, supporting subject-independent classification. The compact single-chirp input and low-dimensional WST representation make the pipeline suitable for edge-AI-enabled IoT health monitoring.
The key advantage of proposed method lies in the IoT radar's ability to perform local, real-time classification without body-worn devices, enabling seamless integration into smart healthcare IoT ecosystems such as intelligent gyms, remote athlete monitoring systems, and ambient-assisted living environments.

Future work will aim to design a multi-modal IoT skin sensing system by collecting concurrent ground-truth measurements of skin temperature, corneometry, sweat electrolyte concentration, and body-mass loss. 
Implementing the entire pipeline on a low-power IoT edge platform (e.g., ARM Cortex-M based microcontroller with an integrated radar front-end) and evaluate its energy consumption, inference latency, and robustness across diverse environmental conditions is another possible direction for future work.

\vspace{-8pt}

 \footnotesize{
 \bibliographystyle{IEEEtran}
 \bibliography{refs}
 }
 
\end{document}